# Ontology Development Kit: a toolkit for building, maintaining, and standardising biomedical ontologies

Nicolas Matentzoglu [1], Damien Goutte-Gattat [2], Shawn Zheng Kai Tan [3], James P. Balhoff [4], Seth Carbon [5], Anita R. Caron [3], William D. Duncan [5, 6], Joe E. Flack [7], Melissa Haendel [8], Nomi L. Harris [5], William R Hogan [6], Charles Tapley Hoyt [9], Rebecca C. Jackson [10], HyeongSik Kim [11], Huseyin Kir [3], Martin Larralde [12], Julie A. McMurry [8], James A. Overton [13], Bjoern Peters [14], Clare Pilgrim [2], Ray Stefancsik [3], Sofia MC Robb [15], Sabrina Toro [8], Nicole A Vasilevsky [8], Ramona Walls [16], Christopher J. Mungall [5], David Osumi-Sutherland [3, *]

[1] Semanticly, Athens, Greece
[2] Department of Physiology, Development and Neuroscience, University of Cambridge, Downing Street, Cambridge CB2 3DY, United Kingdom
[3] European Bioinformatics Institute (EMBL-EBI), Wellcome Genome Campus, Hinxton, Cambridgeshire, CB10 1SD, United Kingdom
[4] RENCI, University of North Carolina, Chapel Hill, NC 27517, USA
[5] Lawrence Berkeley National Laboratory (LBNL), Berkeley, CA 94720, USA
[6] University of Florida, FL 32611, USA
[7] Johns Hopkins University, Baltimore, MD, USA
[8] University of Colorado Anschutz Medical Campus, Aurora, CO 80217, USA
[9] Laboratory of Systems Pharmacology, Harvard Medical School, Boston, MA 02115, USA
[10] Bend Informatics LLC, Oregon, US
[11] Robert Bosch LLC
[12] Structural and Computational Biology Unit, European Molecular Biology Laboratory, Heidelberg, Germany
[13] Knocean Inc., Toronto, Ontario, Canada
[14] La Jolla Institute for Immunology, 9420 Athena Circle, La Jolla, CA 92037, USA
[15] Stowers Institute for Medical Research, Kansas City, MO 64110, USA
[16] Critical Path Institute, 1730 E River Road, Tucson, AZ 85718, USA

*Correspondence: davidos@ebi.ac.uk

## Abstract

Similar to managing software packages, managing the ontology life cycle involves multiple complex workflows such as preparing releases, continuous quality control checking, and dependency management. To manage these processes, a diverse set of tools is required, from command line utilities to powerful ontology engineering environments such as ROBOT. Particularly in the biomedical domain, which has developed a set of highly diverse yet inter-dependent ontologies, standardising release practices and metadata, and establishing shared quality standards, are crucial to enable interoperability. The Ontology Development Kit (ODK) provides a set of standardised, customisable, and automatically executable workflows, and packages all required tooling in a single Docker image. In this paper, we provide an overview of how the ODK works, show how it is used in practice, and describe how we envision it driving standardisation efforts in our community.

# Introduction

In a time of increasing biomedical data output, ontologies have become crucial in research, playing an important role in making data FAIR (Findable, Accessible, Interoperable, and Reusable) (1) by providing standard identifiers (2–5), vocabulary, metadata, and machine-readable axioms (6). Developing high quality and scalable ontologies requires reusing parts of other ontologies, the use of reasoning to automate classification, and extensive quality control (QC) testing. Managing development while following this approach can be a complex process involving tasks such as import management, release file compilation, integration testing, and QC. This difficulty is compounded by the fact that ontologies in the biomedical domain are generally under-resourced, and that biologists, who need to be an integral part of their development, are often not trained in software engineering, and therefore lack exposure to standard best practices for software development. Furthermore, the complexity of the ontology development process is a huge barrier to entry for the community to contribute, limiting the democratisation (and arguably the quality) of these ontologies. Over the years, some ontology communities developed their own workflows for managing the ontology life-cycle using a variety of tools and technical approaches. However, given the complexity of the technologies involved, it is very difficult for even the most experienced ontology pipeline developers to maintain and extend these workflows.

The Open Biomedical Ontologies (OBO) Foundry aims to unify ontologies in the biomedical domain through an evolving set of shared principles governing ontology development, allowing interoperability between ontologies (7,8). However, to effectively achieve this, tools that enable standardisation of these shared principles are needed 1) to support ontology developers to conform to the principles (e.g. through standardised QC as well as standardised release pipelines), and 2) to allow less technical ontology curators to abide by these standards without the need for intensive engineering training.

Ontology engineering is a complex task involving many different workflows such as:

1. Running releases: transforming the ontology through a variety of pipelines involving reasoning, removing redundant content, adding versioning information and exporting to a variety of ontology formats such as OWL, OBO, Turtle, JSON and others.
2. Requesting changes in the form of issues on an issue tracker, and discussing the merit of the proposed changes.
3. Applying changes to the ontology: adding or editing terms, removing logical axioms or changing labels.
4. Reviewing change requests, usually in the form of pull requests.
5. Quality control checking: ensuring that the ontology conforms to a variety of integrity checks, such as logical coherence, label uniqueness and provenance standards.
6. Dependency management: providing methods to import terms from other ontologies, and keep those terms up-to-date in light of changes.
7. Documentation management: providing methods to keep documentation current in the light of changes to the ontology and the ontology workflows.

In this paper, we present the Ontology Development Kit (ODK), a tool for managing the ontology life-cycle. ODK is currently used to maintain more than 70 ontologies, mostly in the biomedical domain, such as the widely used Human Phenotype Ontology (HPO) (9), the Cell ontology (CL) (10,11), Uberon (12), PATO (13), the Brain Data Standards Ontology and Provisional Cell Ontology (14).

The ODK comprises two major components: a set of executable ontology engineering workflows, and a toolbox. It delivers these workflows, which reflect standard best practices recommended by the OBO Foundry, as a customisable git repository setup including all the different files and scripts necessary to run, for example, releases and quality control tests and import terms from other ontologies. The toolbox is delivered as a Docker image and includes all tools necessary to execute these workflows, from command line utilities (sed, git, rsync) to ontology pipeline tools such as ROBOT (15) and dosdp-tools (16).

The ODK simplifies the process of maintaining an ontology, allowing ontology developers to focus on content rather than technical aspects of maintenance. It also allows ontology developers to fully leverage modern "social coding" open source development practices exemplified by many GitHub repositories, such as allowing community contributions via GitHub pull requests, and using cloud-based Continuous Integration (CI) tools to help with QC.

# Motivation

## Sharing best practices

Best practices for ontology engineering evolve over time. For example, it took years of discussions and collective learning to define the OBO Foundry principles, a set of best practices for open, FAIR and interoperable ontology development in the biomedical domain (8), and their refinement is ongoing. While those practices are slowly adopted through a mix of community engagement activities and improved tooling such as the OBO Dashboard (8), the need for extending those practices never stops. For example, there is currently no agreed-upon metadata schema for reflecting contributions to ontology terms, which is critical not only for attribution (grant proposals, individual editors) but more generally important for provenance related questions (Who wrote that definition? Who suggested that term to be added?). To drive this forward, a group of organisations decided to collect this information using a specific property (http://purl.org/dc/terms/contributor) and uniquely identifying IRIs for contributors (such as ORCiDs, Wikidata or ROR identifiers). To ensure this metadata is captured correctly, a schema check needs to be defined (this is typically realised using SPARQL in ODK), and then shared across all participating ontology repositories.

Having a centralised infrastructure like the ODK means that when one ontology faces such an issue, tests (and fixes) can be rolled out via the executable workflows defined by the ODK to all ontologies using it, not just to that particular ontology. This reduces the overhead needed to fix multiple ontologies, and provides a more collaborative environment for problem-solving.

## Standardised repository architecture and release products

Ontologies, even if built with OBO principles in mind, vary in the forms in which they are made available. For example, should the ontology be published with import statements or should the imports be merged in? Should the ontology be published with or without the logical inferences computed by a reasoner? Furthermore, users frequently want ontology files in alternative formats, like RDF/XML, OBO Flatfile format, OBO Graphs JSON or Turtle. Another problem is that the ontology repositories are not usually standardised. Editors who edit more than one ontology have to adjust to the idiosyncrasies of each repository: Which files to edit? How to run tests? How to provide versioned releases? How to add new terms? To address some of these issues, the ODK can automatically generate a standardised file and directory structure that is delivered as a git repository.

Git has become the most widely used version control system in the biomedical ontologies domain. Git repository hosting providers such as GitHub or GitLab have become powerful tools beyond simple version control that cover most aspects of modern software (and ontology) project management, including code review, issue tracking, continuous integration testing, discussions and milestone planning. While not entirely tied in with git and git hosting tools such as GitHub, the ODK is designed in a way that leverages their capabilities. The idea is to encourage best practices promoted by these platforms such as creating and reviewing pull requests that are automatically tested before applying a change, separating source files from release files, and deploying documentation pages side-by-side with the code for ontology development.

The ODK promotes a "convention-over-configuration" model by imposing a standardised repository structure where all files are stored in predictable paths within the repository. This ensures ontology editors are on familiar ground even if they work on multiple different ontologies. The standardised structure includes a strict separation between "source files", which are manipulated by ontology editors and from which the ontology is built (OWL files, files containing DOSDP patterns or SPARQL queries, helper scripts, etc.), and "release files", which result from running the release workflow and are intended for downstream users.

The released version of an ontology can take several forms, depending for example on whether the ontology has been reasoned over or whether it contains imported axioms from foreign ontologies. To facilitate interoperability and modular reuse of ontologies, the ODK defines a few standardised release products, such as the "base" product, which contains only native axioms and the "full" product, which in addition also includes imported axioms and axioms inferred by logical reasoning.

# The ODK Toolbox

The ODK can be divided into two principal architectural components:

1. A toolbox containing everything needed to develop, build, and maintain ontologies, from Unix command line development tools (e.g. rsync, git) to specialised ontology pipeline programs (e.g. ROBOT and fastobo-validator).
2. A set of executable ontology engineering workflows, delivered as a directory of scripts, build rules (e.g. to prepare releases or refresh imports), and source files. These workflows are described in the next section.

The goal of the ODK toolbox is to provide ontology editors with all the tools they need to build, test, and release their ontologies. Tools are chosen for their ability to support the core workflows for managing the ontology life cycle, such as running releases and quality control. A selection of tools (15–21) included in the ODK can be found in Supplementary Table 1. As those tools are very diverse and use different technologies, we cannot merely provide installation instructions that work reliably across the operating systems and computer architectures routinely used by ontology editors. We also lack the resources to provide customised installation packages for all those systems and architectures.

We, therefore, decided to use a Docker image (22) as a software distribution mechanism. Docker is a tool that automates the deployment of applications inside software containers. While it originated on GNU/Linux systems, it is now available on Windows and MacOS as well. The ODK Docker image is based on the Ubuntu base image, in which all the tools listed in Supplementary Table 1 are already installed. All of the core ontology tools and most of the python dependencies are explicitly versioned by the ODK developers, and upgrading them has to be done explicitly, requiring extensive testing. Ontology editors just need to install Docker itself on their platform, and fetch the ODK image from the Docker Hub (https://hub.docker.com/r/obolibrary/odkfull). This saves a great deal of time for both the developers/system administrators and the editors, since the Docker image is effectively a "plug and play" application that can run on any major operating system.

We provide two distinct Docker images. The *ODK-Lite (obolibrary/odklite)* image contains only the minimal set of tools needed by the standard workflows described in the next section. The *ODK-Full (obolibrary/odkfull)* image includes additional tools that an ontology developer may need for some customised, ontology-specific workflows. With one of the ODK images available to the local Docker daemon, ontology editors can invoke any of the provided tools inside the container, without needing to do any additional setup.

# Executable ontology development workflows

The executable ontology engineering workflows are delivered as an ODK-generated Makefile. Targets in that Makefile can roughly be divided into those that provide the recipe for generating a specific file (such as the release file of an ontology) and those that provide simple workflows, such as "clean" to delete temporary files, "prepare_release" to execute the release workflow or "refresh-imports" to update the terms in all imported ontologies. In the rest of this section, we will discuss some of the workflows prevalent in the biomedical ontology community and describe how they are supported by ODK.

## The initialisation and update workflows

The initialisation workflow is performed once in the lifetime of an ontology to create a new ODK setup. Unlike the other workflows, it is launched not from a Makefile but from a small wrapper script that uses the ODK Docker image to:

1. create a new directory which contains all files necessary for editing and managing the new ontology (importantly, this includes the automatically generated Makefile that will pilot all the other standard workflows described in the rest of this section);
2. make a Git repository of the newly created directory;
3. generate an initial release based on the empty ontology.

This initialisation process can be parameterized using either command-line arguments to the wrapper script or a small, YAML-formatted configuration file. Once the repository is set up, it can be pushed to a Git hosting service such as GitHub. The choice of the hosting provider is left to the user, but when GitHub is used, the ODK provides special support for automatically triggering a few server-side workflows (as GitHub Actions) upon certain conditions, e.g. to run a continuous integration test suite whenever a pull request is submitted.

The ODK is continually being updated with new functions, better support, and updated tooling. This allows us to be highly adaptive to the ever-changing landscape of ontology development, and maintain relevance to the community. ODK updates are semi-automated, with a three step process:

1) Update the ODK Docker image
2) Run the ‘update_repo’ command
3) Commit the changes into the ontology repository

We have decided to leave the process of updating the repository to the user rather than folding the workflow into the ODK Docker image itself. This way, the dissemination of new features is a bit slower, but it also gives more control to the ontology engineer to postpone the implementation of potentially breaking changes, such as quality control checks added to the default setup.

## The editors' workflow

Editors frequently change the contents of an ontology by adding or obsoleting terms, revising logical axioms or updating the metadata. While there are many variants of the editors' workflow, i.e. the sequence of actions that lead to the final application of a change, it can be roughly divided into the following steps:

1. The ontology developer (OD) opens the editor's file in their preferred ontology development environment (e.g. Protégé) and makes a change (e.g. adding a term, changing a label). Alternatively, a template file (like a ROBOT template) is edited that first needs to be transformed into OWL.
2. The OD creates a new Git branch locally, commits the change, and opens a pull request on the ontology’s public repository on GitHub.
3. A continuous integration (CI) test suite job configured by the ODK is executed automatically once a pull request is created. This job executes a series of standard and customisable tests, such as looking for unsatisfiable classes, malformed cross-references or missing labels.
4. If the test fails, the developer can inspect the execution log and proceed to fix the problem.
5. Once the test passes, another member of the ontology’s editorial team reviews the change. Depending on the ontology, one or more approvals may be required before a change is merged in, after which the changes will appear in the release products when the release pipeline is run.

The ODK plays two major roles in the workflow, applying changes from templates and coordinating and executing the test suite that ensures that the edit did not “break anything”, i.e. violate one or more of the quality control rules. Templating systems are integrated into the ODK and will be covered in detail in the “Support ODK Features” section later in the manuscript. The test suite provides built-in quality control and continuous integration that can be customised to the individual users' needs.

## Quality control and continuous integration

Even the most experienced ontology curators make mistakes when editing an ontology, from simple ones such as introducing unwanted whitespace in an ontology term label (i.e. a trailing or leading space character), to more complicated errors that lead to unintended logical consequences (e.g. axioms that render a class unsatisfiable, or that cause two distinct classes to be wrongly inferred to be equivalent). To avoid adding such “breaking changes” to the ontology, the ODK uses continuous integration to automatically run quality control tests when a pull request is created or updated. These tests can easily be adopted by other version control providers such as Bitbucket or GitLab, as long as they provide a way to run Docker-based workflows. The QC tests are on the ODK Docker image via a Makefile target, “make test”.

The ODK comes with a wide range of standard QC checks that utilise ROBOT (15), including SPARQL-based validation and logic-based validation. ROBOT incorporates a customisable validation framework for ontologies (called ROBOT report) that performs checks like “illegal trailing whitespace”, “illegal cross-reference syntax”, “missing license” and others. These checks reflect the best practices of the OBO Foundry - if any are not applicable to a specific ontology, they can be skipped, and other checks can be added. Custom SPARQL-based validation provided by the ODK in addition to the standard ROBOT checks follows the same general idea as ROBOT report: an “anti-pattern”, e.g. an undesirable situation like a non-obsolete term without a label, is specified as a SPARQL select query and then executed using ROBOT verify. Logical checks involve running the reasoner using the reason function in ROBOT, and ensuring logical coherency (i.e. the absence of unsatisfiable classes) and the absence of unintended logical equivalencies (i.e. cases where a change to the logical axioms lead to two classes being inferred as logically equivalent that are conceptually distinct). The validation framework is made to be easily customisable. For example, an organisation might have the requirement that new terms are always signed with a (valid) ORCID (23), so they could implement a SPARQL query which looks for terms in the ontologies’ namespace that do not have the respective annotation present and further checks that if the annotation is present the annotation value is a valid ORCID.

## The Release Workflow

Generating release versions of the ontology, including different syntaxes (e.g. JSON, OBO, RDF/XML) and variants (see Section X) is entirely automated as part of an ODK workflow called “prepare release”. To execute this workflow, the ontology developer simply runs a single command on the command line (sh run.sh make prepare_release). The release workflow executes the following steps:

1. Ensure that any automatically generated components (such as portions of the ontology managed as ROBOT or DOSDP templates) are converted into OWL.
2. Ensure that dynamically imported terms are up to date.
3. Generate all release variants, such as the “base”, “full” and “simple” variants using standardised ROBOT pipelines. This includes serialising these release variants into all configured formats such as RDF/XML and OBO format, and adding versioning information.
4. Execute all quality control tests and generate QC reports.

Under the hood, “make prepare_release” builds all release targets (release file variants which are configured as Makefile targets) by loading the editors file and performing (mostly ROBOT-based) transformation pipelines such as merging, reasoning and adding version information. These release targets in turn depend on others, such as up-to-date imports, which are executed as part of the pipeline. After all the release files (also called release assets) are generated, the ontology developer will usually commit the release files to a branch and optionally ask for a review to ensure all the changes to the release files are intended. Depending on how releases are being managed, which differs from ontology to ontology, the last step in the release workflow is to publish the release, which usually involves merging the release to the “main”, or production, branch and publishing a release (e.g. GitHub release). There are some experimental workflows in ODK that automate even this last part of the release process, but in our experience, ontology curators value the opportunity for a “final check” before an ontology release is published.

## Dependency management: Importing and reusing existing ontologies

Ontologies, just like software, can be developed in a modular fashion. Many ontologies make use of an external ontology to provide logical definitions. Previously the ontology community has had a wide range of practices for managing these kinds of inter-ontology dependencies, ranging from copying-and-pasting external terms into an ontology, making duplicative terms in their own ID space, or using the owl:imports mechanism. Using imports is considered best practice, but even here there is a range of different practices. Some ontologies import an external ontology in its entirety, while others import subsets of external ontologies, with a diverse range of methods for creating these subsets. Importing an entire ontology can lead to scalability issues, especially when the external ontology is large (e.g. CHEBI). Importing subsets of external ontologies can also be problematic, since these may be transitively imported by other ontologies. Additionally, external subsets can get stale.

The ODK supports what is considered best practice in OBO, and aims to make it easy for ontology editors to manage imports. An ontology developer can list the external terms they require in a text file managed in git. They can then trigger an ODK workflow that uses ROBOT to generate an “import module” using the appropriate ontology modularisation method, such as syntactic-locality-based module extraction (SLME) (24).

ODK will also take care of making special releases of ontologies that avoid the stale subset problem. A special reusable component release called a base ontology is created. This component includes any terms belonging to the ontology natively and all their axioms, but doesn’t include any of the imported axioms. This base module can then be used as a modular component by other ontologies. The ODK will continue to evolve with best practices in this area.

## Support ODK features

### Using template-based workflows for ontology editing

The ODK supports templating systems such as Dead Simple Ontology Design Patterns (DOSDP) (16) and ROBOT templates (15), which allow ontology content to be curated in the form of spreadsheets, which could include new terms, axioms, or annotations. These spreadsheets are compiled by the ODK into their OWL representation using a simple pattern. Many such patterns have been developed in recent years and are available to be reused. Increasing commitment to patterns will ensure consistent axiomatisation in a scalable manner and increase interoperability and reuse between open ontologies. DOSDP configuration files and templates can be placed in fixed folders in the standard ODK setup, which will then be automatically integrated into the ontology during the build process.

### Auto-generated documentation

Given the complexity of the entire development lifecycle of an ontology, it is important to carefully document all workflows. This documentation includes instructions on how to contribute, how to run a release and how to refresh an import. The ODK repository generation process generates a template for such a documentation system, and auto-generates documentation of the most important ontology workflows tailored to the ontology. For example, rather than providing generic examples of which files to edit during the editors workflow, the specific files used by the ontology (e.g. cl-edit.owl for the Cell Ontology) are mentioned. The documentation system is based on mkdocs (25) and is easily extensible to accommodate the documentation of custom workflows. When using GitHub as the git hosting provider, updates to the documentation can be automatically deployed using GitHub actions. In addition to auto-generated documentation, we have written a number of tutorials on how to set up a new repository or update an existing one (https://oboacademy.github.io/obook/).

## Governance, community requests, quality control, releases

The core ODK team selected and centralised the tools required to optimally support the executable workflows described above. In addition to the core tools required (such as ROBOT, see above) we include a wide range of other tools that are useful for processing of ontologies. The ODK is set up as a community-driven resource in which tool requests and suggestions are encouraged on the issue tracker (https://github.com/INCATools/ontology-development-kit). The possible addition of tools is assessed by the core team. In addition to being open source and free to use, tools should be demonstrably useful for managing some part of the ontology life cycle.

The workflows available in the ODK and reported here were designed with OBO principles in mind. We are also working with members of the FAIR semantics (26) community to incorporate practices beyond the OBO standards (for example by including the owl:versionInfo property as part of the ontology metadata).

Any change to the ODK Docker image, in particular upgrading the tools in the ODK toolbox, is carefully evaluated by a large set of integration tests which are executed every time a new feature or upgrade is proposed in a pull request, and just before every ODK release. These tests include building and running a large variety of different configurations of ontologies, as well as testing the integrity of some of the support tools directly using a shell script which is executed as part of the build process.

The ODK is scheduled for a new release every 3 months, but occasionally, an additional release is required to update a tool that has implemented a critical bug fix. Overall, more than 30 releases were created in the last 3 years. All new tool additions are documented in the changelog of each release. Users can subscribe to be informed about new releases through GitHub's "Notification" feature.

# Use across OBO and beyond

The ODK is designed to allow use at different levels of "buy in". Some ontologies are entirely automated using ODK. Others have partial adoption, such as using the ODK container to run custom-built workflows/checks.

We evaluated the use of ODK using a number of methods, but we do not claim to provide a full account of the usage, just a lower bound. We first gathered some of the ontologies that we already knew were using ODK. We also performed GitHub searches to see which additional repositories were using ODK (using search terms like "ontology development kit" or "ontology starter kit"). Lastly, we performed an informal user survey targeting the OBO ontology community (distributed via the obo-discuss mailing list), which got 35 responses, 23 of which had used the ODK. Among the surveyed ontologies, 61.3% were in the OBO foundry.

The ODK Docker image was pulled 74611 times from Docker Hub at the time of this writing (22.06.2022). Note that this merely establishes a lower bound, as the Docker image is cached locally, and therefore does not need to be pulled more than once per 3 months by most developers. Most of these pulls are probably generated by various continuous integration tools, which are difficult to differentiate from human users. A table of all ODK-based ontology repositories and a summary of ODK versions used can be found in the supplemental materials (Supplementary Table 2, Supplementary Figure 3), and the results are summarised in Figure 1.

# Related work

The ODK is not the only ontology development toolkit used by ontology developers. In this section, we will highlight some popular tools that are currently being used and how the ODK compares with them.

Firstly, arguably the most popular ontology development tool is Protégé. Protégé (27) is a visual ontology editor that is free and open source. Protégé has a few key features that make it a highly useful tool. Firstly, it has a graphical user interface that is relatively easy to use and navigate, making it highly accessible to non-technical users. It also has a library of plugins that can further extend its functionality, and since it is open source, it is possible to develop new custom plugins. The ODK was developed not as a replacement for Protégé, but rather a complement to it. The ODK is not aimed at the actual process of ontology editing, and therefore lacks any support for data entry (such as visual editors). Protégé also has a lightweight cloud-based version (webProtégé) (28) that can be accessed either through their or a hosted server, allowing better support for simultaneous collaborative editing of OWL ontologies. The ODK does not currently integrate with WebProtégé editing workflows, but exploring this integration is on our list for the future.

The OntoAnimal set of tools in conjunction with the eXtensible ontology development (XOD) principles (29) is another toolkit that is widely used across OBO. Our work builds upon the general XOD principles, such as “re-use” and “bulk-import”, and we hope to reconcile some of our workflows over the coming years. There are some differences to overcome. Firstly, the OntoAnimals toolkit does not provide standardised executable workflows (releases workflows, dynamic imports for an entire project), which is at the heart of the ODK design. Some of the OntoAnimals best practices could therefore simply be integrated into the ODK workflow system. Additionally, the preferred way to extract modules in the wider OntoAnimals culture is to use MIREOT modules through a web service. This is different from the ODK culture in two ways: 1) one key design decision of ODK is to go straight to a source ontology from which to import terms, download it and extract a module from the file, while the OntoAnimals approach tends to use the OntoFox web-service to extract terms. We prefer our approach because of the lack of intermediary (the imports are less often out-of-date when not using an intermediary) and less pronounced network dependency (the web service might become unavailable, or a network could be down). For example, we can extract a module from a “mirror” over and over without downloading the mirror again. However there are downsides to our approach: the often quite large ontologies require large amounts of resources (memory, CPU). The best approach has not been definitively determined, but it should be mentioned that OntoAnimals module extraction workflows can be easily integrated into the ODK through its customisation features. 2) A second conceptual “conflict” between OntoFox and ODK is the use of MIREOT modules vs. SLME (syntactic-locality based modules). MIREOT modules are often more intuitive for downstream consumers of the ontology, because they only expose the terms that are deliberately imported by the ontology developer, while SLME based modules import all terms that are necessary to preserve the semantics of the the source ontology (which is usually a lot). While we believe that SLME modules are necessary to ensure that our ontology is logically compatible with our dependencies at development time, we believe that neither MIREOT nor SLME modules are sufficient for downstream users. We are currently designing a module extraction algorithm that not only traverses the subclass hierarchy (like MIREOT does) but also takes into account other kinds of relationships like mereological ones (part of, has part) which are, for biological use cases, equally important (often more important).

The NeOn Toolkit (30) is an open-source ontology engineering environment, built on the code-base of OntoStudio (31) (formally known as OntoEdit (32)), that aims to provide comprehensive support for the entire engineering lifecycle by providing an extensive set of plug-ins. It organises projects in a workspace; each project can contain multiple ontologies that can be edited in a graphical user interface, which requires more knowledge than Protégé as much of the interface is pretty technical. A major advantage of the NeOn Toolkit is an open platform to which anybody can contribute, hence allowing for a wide variety of plugins developed by the community.

WebODE (33) is an extensible ontology-engineering suite developed over 16 years ago that aimed to be a common workbench for ontology development and management, middleware services (such as access and query services), and ontology-based application development. WebODE has many functionalities including a simple user interface, complete consistency checks, edition through a form-based user interface or a graphical editor, and term import. However, WebODE support was discontinued in 2006, and while it remains open-source, no support will be provided when problems are encountered.

# Limitations

The dependency on Docker is one of the most significant limitations of the ODK architecture. While Docker itself is not needed to run a Docker container (Docker containers can be run with other container systems such as singularity, which are also widely supported), the non-availability of Docker for some of our users with (a) older Windows machines or (b) no access to admin privileges on their work machines has been challenging. To mitigate this, we have worked on integrating many of the workflows as GitHub actions. For example, rather than running the command to build and deploy the documentation manually, a GitHub action can be launched to do this. The problem with doing the same for other key workflows like ontology import management or releases is that GitHub actions are limited in how much memory they can use (7GB at the time of writing), which is often exceeded, especially for larger ontologies.

# Discussion and conclusions

Creating and managing biomedical ontologies are complex tasks that require deep technical knowledge. Developing an ontology that is standardised to other ontologies and reuses them compounds this difficulty. This is often prohibitively difficult for many biologists and domain experts whose input is needed to make biomedical ontologies accurate and useful. The ODK provides tools and features that allow non-experts to build and edit ontologies with minimal training. The ODK helps ontologies conform to basic standards and sets users up with structures and documentation for good ontology engineering practices. The import management system allows non-technical users to reuse existing ontologies, which would otherwise be incredibly complex. The ODK's built-in tools, such as templating systems, further enhance the users' toolbelt, allowing highly powerful automation with minimal technical knowledge. Overall the simplification achieved through the ODK allows users and developers to focus on the content while standardising good practices and democratising ontology development.

In a future release of the ODK, we plan to reconcile some of its workflows with other existing frameworks such as OntoAnimals, in particular OntoFox, and work with their developers towards a common solution for, at least, biomedical ontologies. Furthermore, there is a great need for better module extraction techniques for downstream usages as both SLME and MIREOT (the most prevalent approaches) fall short in various ways. Lastly, the ODK does not currently prevent bad ontology modelling - we hope to be able to make stronger use of design pattern-based validation and advanced semantic validation techniques such as SHACL (34), SHEX (35,36) or LinkML (37) to further reduce the potential for human error.

We have already observed significantly lower error rates in many of the ontologies that use the ODK, thanks to the ability of the automated testing system provided by the ODK to catch errors early on. We hope to be able to roll out ever more useful tests to ever wider circles of ontologies to contribute to a community-wide increase in ontology quality. Our update system allows us to rapidly roll out new features, such as new quality tests and improved pipelines, to all our users by pulling the new Docker image from Docker Hub and running the "update repo" workflow. Lastly, we seek to harmonise the representation of ontology release files through the use of standard release workflows which result in standard release serialisations (RDF/XML, OBO Flat File, OBO Graphs JSON) and metadata (version IRIs, licence information and more) to make ontologies more FAIR and interoperable.

# Funding

This work was supported by the Office of the Director, National Institutes of Health (R24-OD011883); National Human Genome Research Institute “Phenomics First” (7RM1HG010860-02); NHLBI 5U01HG009453-03; UK Biotechnology and Biological Sciences Research Council / US National Science Foundation Directorate of Biological Sciences (BBSRC-NSF/BIO BB/T014008/1); Director, Office of Science, Office of Basic Energy Sciences, of the US Department of Energy [DE-AC0205CH11231 to C.J.M.]; and the National Institutes of Mental Health (1RF1MH123220-01).

# Supplementary Materials

**Supplementary Table 1.** A selection of tools included in the ODK, their purposes, and the programming languages they use.

| ODK Component | Purpose | Language | In ODK-Lite? | GitHub Repository |
|---|---|---|---|---|
| ROBOT (15) | Primary multi-purpose ontology manipulation tool | Java | Yes | https://github.com/ontodev/robot |
| OWLTools (17) | Former multi-purpose ontology manipulation tool | Java | Yes | https://github.com/owlcollab/owltools |
| DOSDPTools (16) | Work with DOSDP design patterns | Java | Yes | https://github.com/INCATools/dosdp-tools |
| Apache Jena | RDF graphs manipulation | Java | No | https://github.com/apache/jena |
| Konclude (18) | Reasoner | C++ | No | https://github.com/konclude/Konclude |
| Soufflé (19) | Logic programming language | C++ | No | https://github.com/souffle-lang/souffle |
| SPARQLProg | SPARQL query engine | Prolog | No | https://github.com/cmungall/sparqlprog |
| Fastobo | OBO format parser | Python/Rust | No | https://github.com/fastobo/fastobo |
| SOak (20) | Manipulating ontologies with Python | Python | No | https://github.com/INCATools/ontology-access-kit |
| Sssom (21) | Working with SSSOM mapping files | Python | No | https://github.com/mapping-commons/sssom |
| git, make, etc. | General development tools | C | Yes | |
| rsync, wget, etc. | Network tools | C | Yes | |

**Supplementary Table 2.** Ontology repositories known to make use of ODK. The OBO column indicates whether the ontology is a member of the OBO Foundry.

| Repository | Ontology ID | ODK Version | Date checked | OBO? |
|---|---|---|---|---|
| https://github.com/Aur-Int/agro | agro | 1.2.32 | 2022.01.24 | |
| https://github.com/berkeleybop/artificial-intelligence-ontology | aio | 1.2.29 | 2022.06.02 | |
| https://github.com/BCODMO/bcodmont/ | bcodmont | 1.2.23 | 2022.03.09 | |
| https://github.com/obophenotype/brain_data_standards_ontologies | bdso | 1.2.32 | 2022.01.24 | |
| https://github.com/berkeleybop/bero | bero | 1.3.0 | 2022.05.26 | |
| https://github.com/hubmapconsortium/ccf-ontology | ccf | 1.2.29 | 2022.01.24 | |
| https://github.com/cmconnor/childwelfare | childwelfare | 1.2.29 | 2022.01.24 | |
| https://github.com/EBISPOT/CMPO | CMPO | pre-1.2 | 2022.01.24 | |
| https://github.com/data2health/credit-ontology | credit | pre-1.2 | 2022.01.24 | |
| https://github.com/turbomam/deep-learning-ontology | deep-learning-ontology | 1.2.29 | 2022.01.24 | |
| https://github.com/Knowledge-Graph-Hub/kg-obo | kg-obo | image | 2022.03.09 | |
| https://github.com/monarch-initiative/glyco-phenotype-ontology | mgpo | pre-1.2 | 2022.01.26 | |
| https://github.com/monarch-initiative/monochrom | monochrom | 1.2.29 | 2022.01.24 | |
| https://github.com/Buffalo-Ontology-Group/MRI_Ontology | mri | 1.2.27 | 2022.01.24 | |
| https://github.com/microbiomedata/nmdc-ontology | nmdco | 1.2.27 | 2022.01.24 | |
| https://github.com/ufbmi/ORCS/ | orcs | 1.2.29 | 2022.01.24 | |
| https://github.com/monarch-initiative/phenio | phenio | 1.3.0 | 2022.05.27 | |
| https://github.com/hurwitzlab/planet-microbe-ontology | pmo | pre-1.2 | 2022.03.09 | |
| https://github.com/EBISPOT/scatlas_ontology | scao | 1.2.25 | 2022.03.06 | |
| https://github.com/NASA-IMPACT/sddo | sddo | 1.2.29 | 2022.01.24 | |
| https://github.com/AgriculturalSemantics/SEOnt | seont | 1.2.22 | 2022.06.14 | |
| https://github.com/nleguillarme/soil_food_web_ontology | sfwo | 1.2.23 | 2022.03.09 | |
| https://github.com/UA-SRC-data/srpdio | srpdio | 1.2.32 | 2022.01.24 | |
| https://github.com/BICCN/TMN | tmn | 1.2.30 | 2022.01.24 | |
| https://github.com/PennTURBO/turbo-ontology | turbo | 1.2.32 | 2022.01.24 | |
| https://github.com/monarch-initiative/vertebrate-breed-ontology | ubo | 1.2.31 | 2022.01.26 | |
| https://github.com/VirtualFlyBrain/vfb-driver-ontology | VFB_drivers | 1.3.0 | 2022.03.06 | |
| https://github.com/VirtualFlyBrain/vfb-scRNAseq-ontology | VFB_scRNAseq | 1.3.0 | 2022.05.25 | |
| https://github.com/VirtualFlyBrain/vfb-extension-ontology | vfbext | 1.2.32 | 2022.01.24 | |
| https://github.com/krishnato/agro/tree/ODK-rebuild | agro | 1.2.32 | 2022.03.09 | Y |
| https://github.com/insect-morphology/aism | aism | 1.2.31 | 2022.01.24 | Y |
| https://github.com/BiodiversityOntologies/bco | bco | 1.2.30 | 2022.01.24 | Y |
| https://github.com/obophenotype/biological-spatial-ontology | bspo | 1.3.0 | 2022.02.28 | Y |
| https://github.com/BRENDA-Enzymes/BTO | bto | 1.2.19 | 2022.01.26 | Y |
| https://github.com/obophenotype/caro | caro | 1.3.0 | 2022.02.28 | Y |
| https://github.com/Southern-Cross-Plant-Science/cdno/ | cdno | 1.2.22 | 2022.03.09 | Y |
| https://github.com/obophenotype/cell-ontology | cl | 1.3.0 | 2022.03.09 | Y |
| https://github.com/luis-gonzalez-m/Collembola | clao | 1.2.29 | 2022.01.24 | Y |
| https://github.com/OBOFoundry/COB | cob | 1.2.30 | 2022.01.24 | Y |
| https://github.com/insect-morphology/colao | colao | 1.2.31 | 2022.01.24 | Y |
| https://github.com/obophenotype/dicty-phenotype-ontology | ddpheno | 1.2.33 | 2022.01.24 | Y |
| https://github.com/FlyBase/drosophila-phenotype-ontology | dpo | 1.2.32 | 2022.01.24 | Y |
| https://github.com/ufbmi/dron | dron | 1.2.27 | 2022.01.24 | Y |
| https://github.com/EcologicalSemantics/ecocore | ecocore | 1.3.0 | 2022.01.24 | Y |
| https://github.com/EnvironmentOntology/environmental-exposure-ontology | ecto | 1.2.30 | 2022.01.24 | Y |
| https://github.com/EnvironmentOntology/envo | envo | pre-1.2 | 2022.01.24 | Y |
| https://github.com/FlyBase/drosophila-anatomy-developmental-ontology | fbbt | 1.2.32 | 2022.01.24 | Y |
| https://github.com/FlyBase/flybase-controlled-vocabulary | fbcv | 1.2.31 | 2022.01.24 | Y |
| https://github.com/FlyBase/drosophila-developmental-ontology | fbdv | 1.2.31 | 2022.01.24 | Y |
| https://github.com/foodOntology/foodon/ | foodon | pre-1.2 | 2022.03.09 | Y |
| https://github.com/futres/fovt | fovt | 1.2.32 | 2022.01.24 | Y |
| https://github.com/pombase/fypo | fypo | 1.2.30 | 2022.01.24 | Y |
| https://github.com/GenEpiO/genepio/ | genepio | pre-1.2 | 2022.03.09 | Y |
| https://github.com/monarch-initiative/GENO-ontology | geno | 1.2.22 | 2022.01.26 | Y |
| https://github.com/obophenotype/human-phenotype-ontology | hp | 1.2.32 | 2022.01.24 | Y |
| https://github.com/insect-morphology/lepao | lepao | 1.2.30 | 2022.01.24 | Y |
| https://github.com/monarch-initiative/MAxO | maxo | 1.2.26 | 2022.01.24 | Y |
| https://github.com/monarch-initiative/mondo | mondo | pre-1.2 | 2022.01.24 | Y |
| https://github.com/obophenotype/mammalian-phenotype-ontology | mp | 1.2.31 | 2022.01.24 | Y |
| https://github.com/obo-behavior/behavior-ontology | nbo | 1.2.26 | 2022.01.24 | Y |
| https://github.com/obophenotype/bio-attribute-ontology | oba | 1.2.30 | 2022.01.24 | Y |
| https://github.com/ufbmi/OMRSE | omrse | 1.2.27 | 2022.01.24 | Y |
| https://github.com/pato-ontology/pato | pato | 1.3.0 | 2022.02.28 | Y |
| https://github.com/obophenotype/provisional_cell_ontology | pcl | 1.3.0 | 2022.02.28 | Y |
| https://github.com/PHI-base/phipo | phipo | 1.2.28 | 2022.01.24 | Y |
| https://github.com/obophenotype/planaria-ontology | plana | 1.2.30 | 2022.01.24 | Y |
| https://github.com/obophenotype/planarian-phenotype-ontology | planp | 1.2.20 | 2022.01.24 | Y |
| https://github.com/obophenotype/uberon | uberon | 1.3.0 | 2022.03.01 | Y |
| https://github.com/obophenotype/c-elegans-gross-anatomy-ontology | wbbt | 1.2.22 | 2022.01.24 | Y |
| https://github.com/obophenotype/c-elegans-development-ontology | wbls | 1.2.31 | 2022.03.09 | Y |
| https://github.com/obophenotype/c-elegans-phenotype-ontology | wbphenotype | 1.2.32 | 2022.01.24 | Y |
| https://github.com/obophenotype/xenopus-phenotype-ontology | xpo | 1.2.27 | 2022.01.24 | Y |
| https://github.com/ybradford/zebrafish-experimental-conditions-ontology | zeco | 1.2.23 | 2022.01.24 | Y |
| https://github.com/cerivs/zebrafish-anatomical-ontology | zfa | 1.2.31 | 2022.01.24 | Y |
| https://github.com/obophenotype/zebrafish-phenotype-ontology | zp | 1.2.31 | 2022.01.24 | Y |

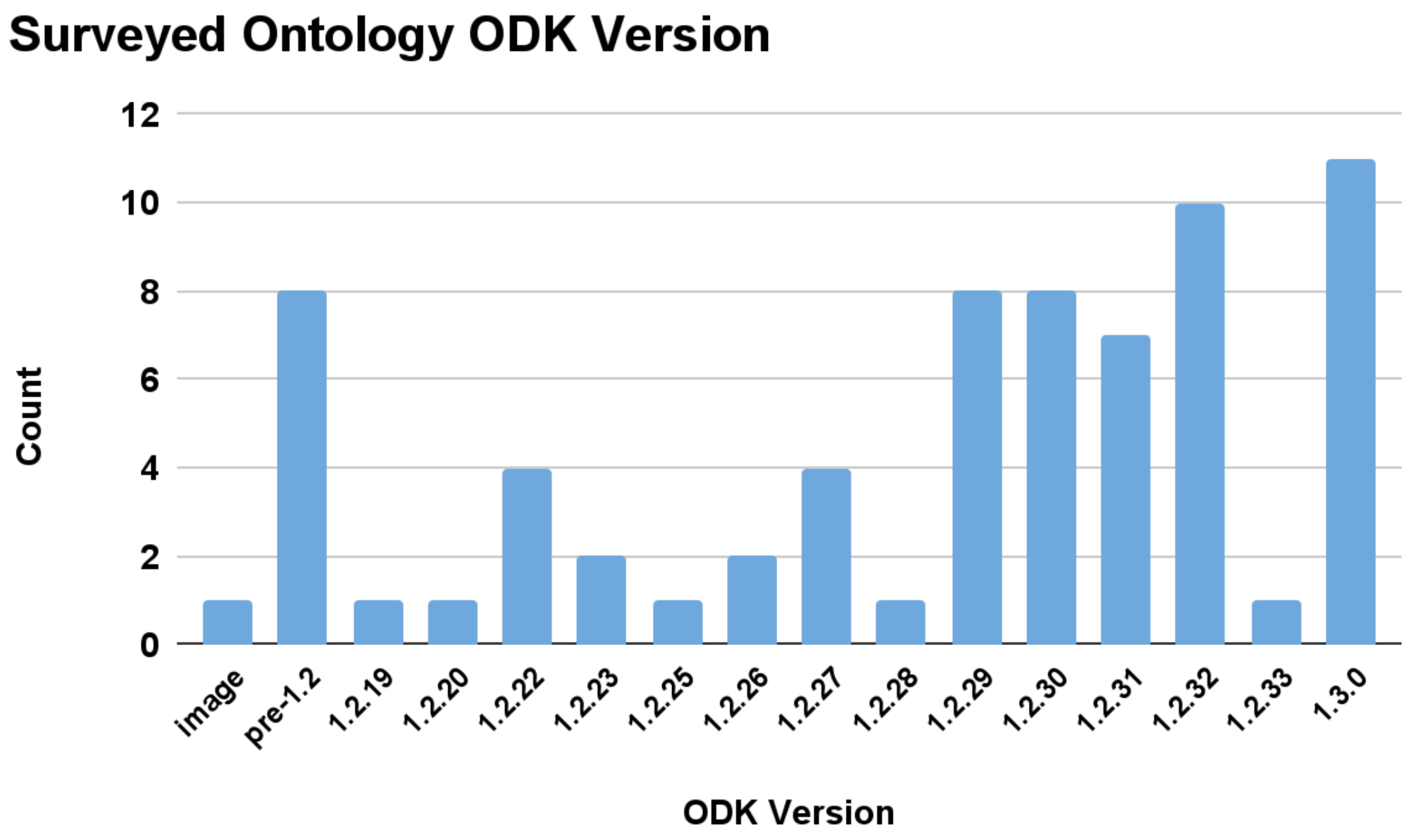

**Supplementary Figure 3.** Results of informal survey of ODK based ontologies showing the versions of ODK used as of time of data collection.